# Determination of the National Highpoint of Botswana using GNSS

Eric Gilbertson[a]*, Mooketsi Segobye[b], Yashon Ouma [b], Boipuso Nkwae [b]

*Mechanical Engineering, Seattle University, Seattle, USA; [b]Civil Engineering, Botswana University, Gaborone, Botswana*

*gilberte@seattleu.edu

# Determination of the National Highpoint of Botswana using GNSS

Botswana has not previously been surveyed with sufficient accuracy to determine the highest peak in the country. Otse Hill and Monalanong Hill have been identified as the highest peaks, but there was uncertainty on which is highest. For this study, ground surveys were conducted on each of these peaks using an Abney level to identify the peak location and a GNSS unit to measure the elevation of each hill with sub-meter vertical accuracy. Monalanong Hill was measured to be 1.87m +/-0.02m taller than Otse hill, with an elevation of 1492.12m +/-0.01m (orthometric height, EGM2008 geoid, 95% confidence interval) above mean sea level (MSL). Otse Hill was measured to be 1490.25m +/-0.01m. Thus, Monalanong Hill is the highest hill in Botswana.

**Introduction**

The location of the highest peak in Botswana is geographically significant. The location of the highest point can be a tourist attraction and therefore uplift the economy of the local community. However, before this study, the elevation and location of the highest peak was not known with certainty. Previously, conventional survey methods and satellite-based measurements have been conducted.

Otse Hill was measured by a trigonometric ground survey in 1999 by the Botswana Department of Surveys and Mapping (DSM) (Botswana 1999). This measured Otse Hill 1491m, but Monalanong Hill (Table 1) was not measured. The purpose of the survey was to place a geodetic station. Hence, they did not observe the highest peak on the plateau, but a point suitable for geodetic control. In the current survey an Abney level was used to identify the high point on Otse hill.

In February 2000, the satellite-based Shuttle Radar Topography Mission (SRTM)

collected elevation data at discrete points around the world, including in Botswana, using radar interferometry. The data was processed to create earth's surface topography at a spatial resolution of 1 arcsecond (approximately 30m) as of September 2014. According to Farr et.al and Smith 2023 et. al, elevations from SRTM have a vertical accuracy +/-16m. Elevations of locations between grid points of the SRTM Digital Elevation Model (DEM) are estimated. However, the error bounds of elevations of locations between measured points is unknown and can potentially be higher than +/-16m, especially for sharp peaks (Sandip 2013).

Digital elevation models based on SRTM data include Google Earth (Google Earth 2025), Google Terrain (Google Terrain 2025), Topographic-map (Yamazaki 2017), Floodmap (Floodmap 2025), and Gaia (Gaia 2025). All of these models identified the two highest peaks in the country as Otse Hill (location 25°0'27.6732" S, 25°25'56.0376" E) and Monalanong Hill (location 24°50'25.5732" S, 25°39'54.8244" E) (Fig 1). All other peaks in Botswana were lower and outside the error bounds of the elevation of these two peaks. The peaks have the same elevation within the error bounds of the measurements, and thus satellite-based measurements are not sufficient to distinguish which peak is highest (Table 1).

Thus, based on all existing measurements, it was unknown which of the two peaks is highest. Since the survey by DSM was not for determining the highest point and satellite-based methods only provide estimated elevations, it is imperative to carry out a ground survey to measure accurate elevations. GNSS offers the most efficient and accurate way to collect data. However, GNSS heights are ellipsoidal heights, which are simply geometric heights with no physical attributes. Hence the observed ellipsoidal heights will be converted to orthometric height using global geopotential models.

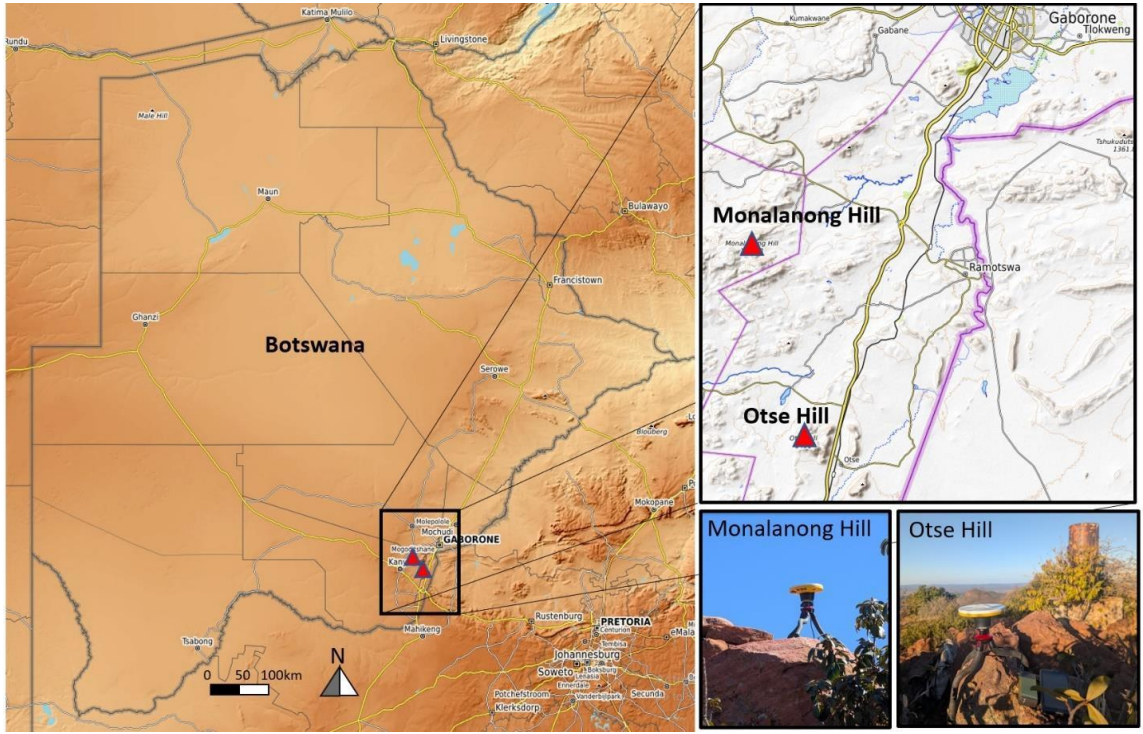

Figure 1. Location of Otse Hill and Monalanong Hill in Botswana. Basemap from Opentopo (Opentopo 2025).

Table 1: Elevations (m) for each peak from different sources (EGM96 geoid).

| Monalanong | -- | 1492 | 1491 | 1494 | 1495 | 1480-1500 |
|---|---|---|---|---|---|---|
| Otse | 1491 | 1490 | 1489 | 1487 | 1486 | 1480-1500 |
| Data type | Map | DEM | DEM | DEM | DEM | DEM |
| Spatial resolution (m) | -- | 30 | 30 | 30 | 30 | 30 |
| Source | 1:50k DSM Map | Google Earth | Topographic-map | Floodmap | Gaia | GoogleTerrain |

**Materials and Methods**

This study was based in South-eastern Botswana, where the two hills are located. Monalanong is in Mogonye village (Southern District), while Otse hill is in the village of Otse in the Southeast district, about 20 km to the southeast of Monalanong (Fig 1). A 10 arcminute 5x Sokkia Abney level was used to identify the peak on the plateau of each hill. This was to ensure that GNSS observations are on the true summit. Once the peak was established, a Trimble DA2 differential GNSS unit capable of sub-meter vertical accuracy was used to observe the high point. The DA2 unit is a portable and lightweight (330 g) instrument, that can connect wirelessly to an android™ and iOS devices, making it ideal for this survey. The multi-frequency DA2 uses satellite-based augmentation services (SBAS) for differential corrections. It is also capable of recording static data, allowing for post-processing and using stable GNSS satellite ephemeris.

The first point to be observed was Monalanong Hill, on August 7, 2025. The location was first approximately identified by satellite DEMs on a broad plateau. Once on the plateau, an Abney level was used to identify the highest rock in this vicinity. This was a rock pillar approximately 2m x 1m x 1m sticking up from the plateau (Fig 2).

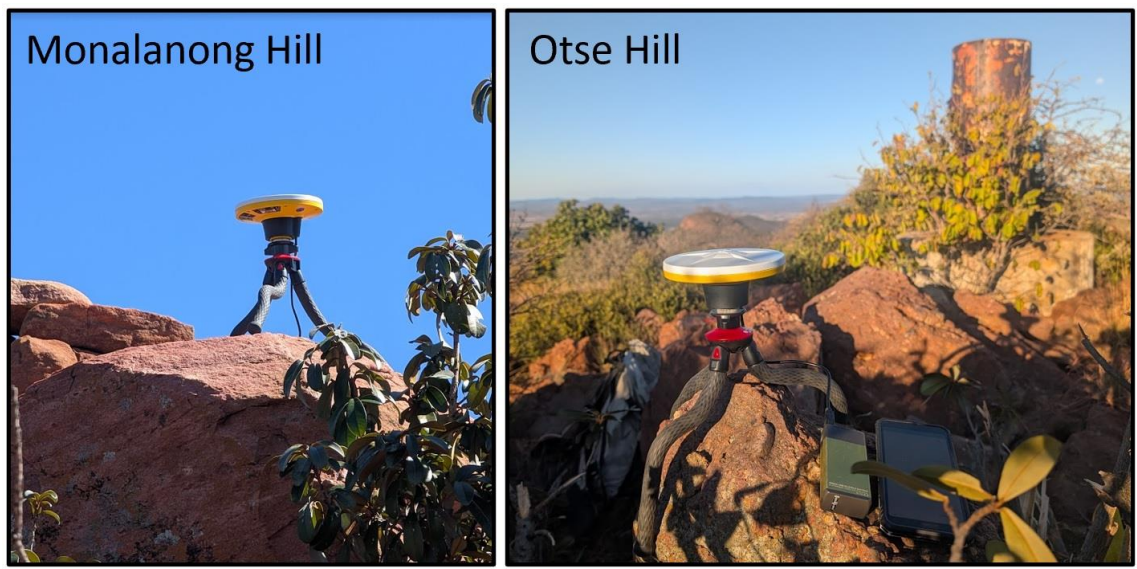

Figure 2. Trimble DA2 setup on Monalanong Hill and Otse hill.

We mounted the DA2 on a flexible-leg tripod on the highest point and logged data for one hour one minute at 1 Hz. The antenna height was measured multiple times and the average height recorded as 0.16m above ground. A total of 3708 satellite observations were made (Table 2).

Table 2: Duration and number of observations for each location.

|            | Start - finish        | No of observations |
|------------|----------------------|--------------------|
| Monalanong | 09:33:33 -10:35:20   | 3708               |
| Otse       | 14:49:13 – 15:57:58  | 4126               |

Later the same day, we ascended Otse Hill to the concrete monument at the summit. The Abney level was used again to identify the highest rock in the summit vicinity, and we mounted the DA2 on the rock on the tripod (Fig 2). The antenna height was 0.08m, and we logged data for over one hour. The survey started at 14:49:13 and finished at 15:57:58 making 4126 observations. In all for each setup multiple frequencies were received, with GPS and Beidou having 12 observation types, GLONASS and Galileo 8 and IRS with 4.

*Processing*

Three processing strategies were used including TrimbleRTX, CSRS-PPP, and PRIDE PPP-AR. All the strategies are based on precise point positioning (PPP). PPP is a GNSS technique that determines the position of a single receiver with high accuracy by using precise satellite orbits and clock corrections. With PPP there is no need for a reference station, making it appropriate for remote areas and reducing the need for a base station, as is the case for high-accuracy relative positioning. TrimbleRTX is Trimble's PPP service that provides corrections by using Trimble's globally located reference stations. TrimbleRTX can provide both real-time and post-processed solutions. CSRS-PPP is a software-as-a-service (SaaS) post-processing product by the Canadian Geodetic Service.

Currently, CSRS-PPP can process GPS, GLONASS and Galileo data. PRIDE PPP-AR is an open-source software by the PRIDELab, Wuhan University. Unlike the other products it is not web-based. PPP-AR improves PPP by recovering the integer property of ambiguities (ambiguity resolution, AR), hence taking less time to converge and achieving high accuracy (Geng 2016 et. Al).

Processed GNSS data will yield horizontal and vertical position, with the vertical position being the ellipsoidal height. This height needs to be converted to an orthometric height, which will be done by using the relationship between reference ellipsoid and geoid as given in equation 1.

$$H=h-N \qquad 1$$

Where *H* is the orthometric height, *h* is the ellipsoidal height and *N* the geoidal height or geoid undulation (Fig 3). *N* is the height difference between the reference ellipsoid and geoid at a specific point on Earth. *N* can be derived from local geoid models, however where there are no local geoid models, global geopotential models can be used. In this case, *N* will be extracted from the geoid height calculator (UNAVACO) which uses EGM2008.

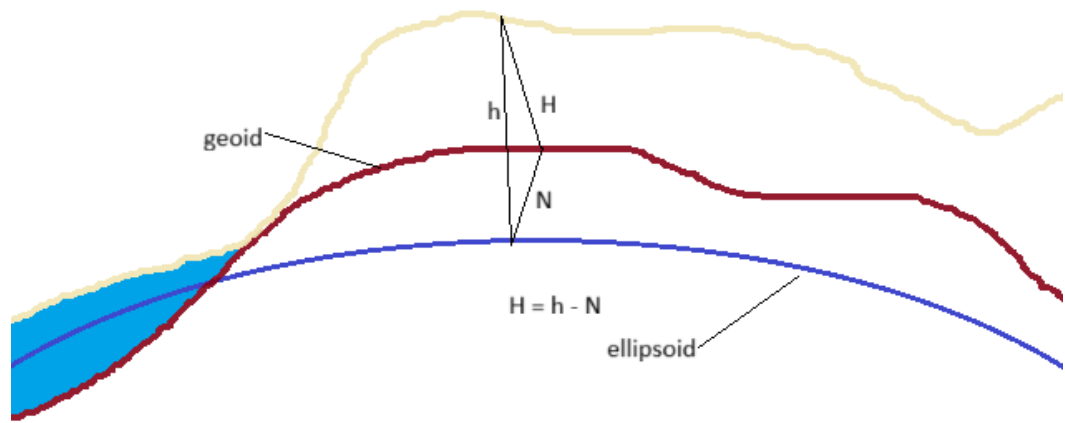

Figure 3. Relationship between ellipsoidal height, geoidal height and orthometric height.

**Results**

Elevations are reported as orthometric height using the EGM2008 geoid (Pavlis 2012), because this is the most accurate geoid for this area of Botswana. We converted ellipsoidal heights to orthometric heights using the UNAVCO/NSF GAGE tool (UNAVCO 2025). All error bounds will be reported as 95% confidence intervals.

*GNSS processing*

We processed data with PRIDE-PPPAR (Geng 2023), TrimbleRTX (Trimble 2025), and CSRS-PPP (Tétreault 2005) accounting for antenna rod heights. Results from CSRS-PPP processing (Fig 4) can be seen converging to final values over the one-hour measurement. However, the convergence was slow. It took close to 20 minutes because ambiguity resolution is not used. The static solutions for these measurements resulted in errors in elevation of +/-0.051m for Monalanong Hill and +/-0.052m for Otse Hill, with Monalanong Hill 1.99m +/-0.07m taller than Otse Hill.

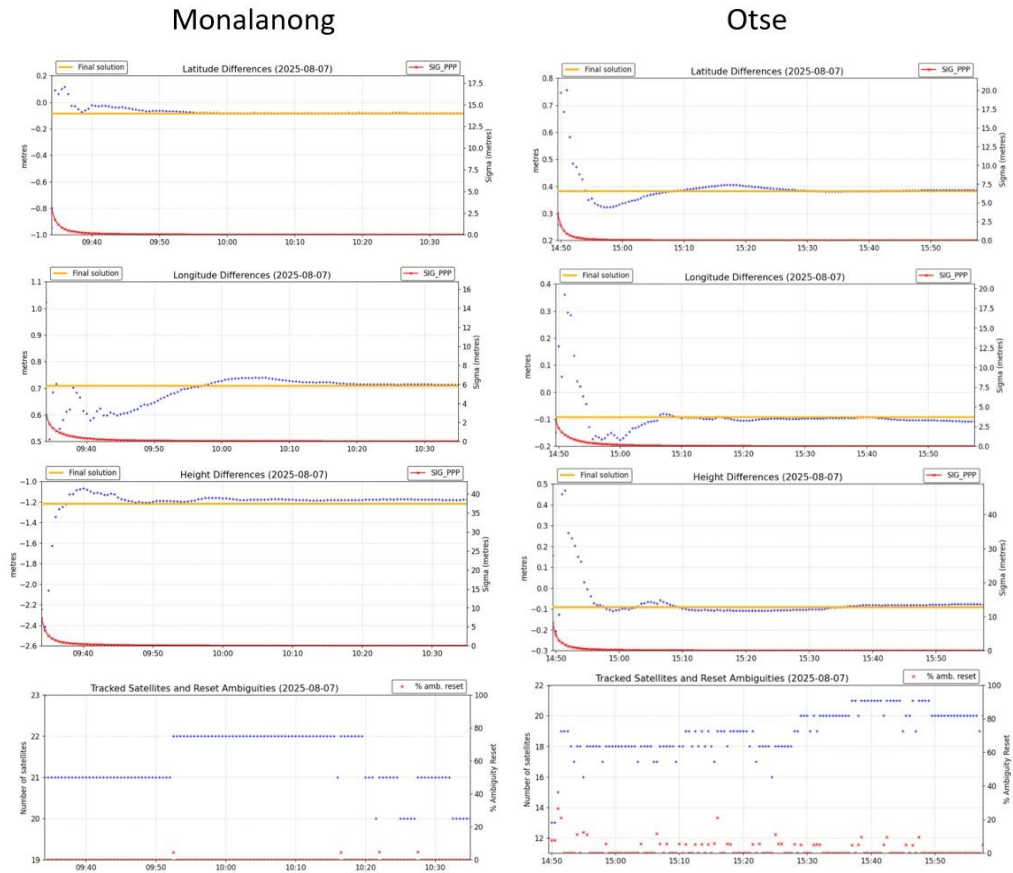

Figure 4: Convergence of solutions using CSRS-PPP.

Results from TrimbleRTX static processing resulted in errors in elevation of +/-0.092m for Otse Hill and +/-0.106m for Monalanong Hill. Monalanong Hill was found to be 2.07m +/- 0.14m taller than Otse Hill.

We processed using PRIDE-PPPAR to first find kinematic solutions (Fig 5). This showed vertical errors for Monalanong Hill of +/-0.0137m and for Otse Hill +/-0.0105m.

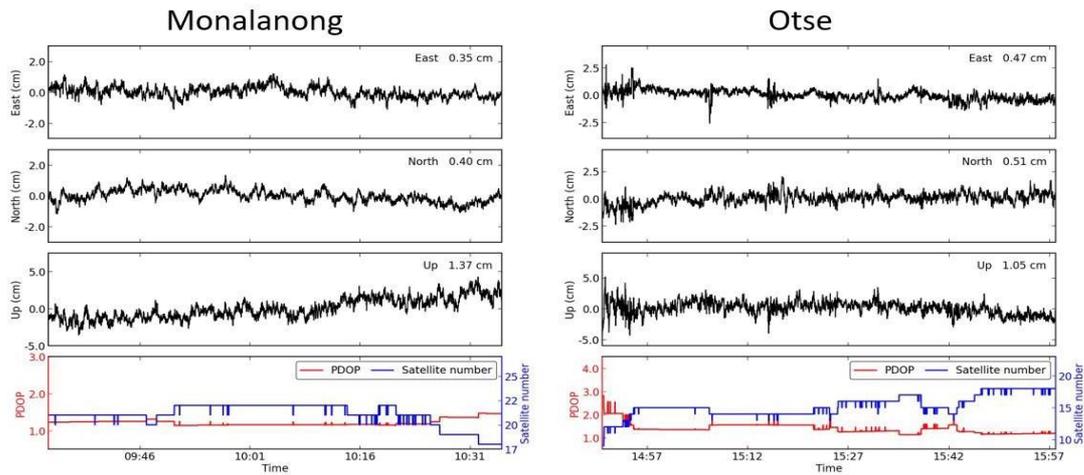

Figure 5: Convergence of solutions using PRIDE –PPP-AR.

We next processed the data with PRIDE –PPP-AR to find a static solution for each location, which gave vertical errors of +/-0.00054m for Monalanong Hill and +/-0.00071m for Otse Hill. PRIDE PPP-AR solution converged faster when compared to CSRS-PPP. Static solutions generally have smaller errors than kinematic solutions, and final results will thus be reported as the static results. All methods found the ellipsoidal height of Monalanong Hill to be 1.368 m higher than Otse Hill (Table 3).

Table 3. GNSS post-processing results.

|  | Coordinate | Trimble RTX | CSRS-PPP | PRIDE PPP-AR |
|---|---|---|---|---|
| **Monalanong** | Latitude | 24°50'25.57344" | 24°50'25.57335" | 24°50'25.57327" |
|  | Longitude | 25°39'54.82596" | 25°39'54.82858" | 25°39'54.82874" |
|  | Ellipsoidal height (m) | 1514.994 ± 0.053 m | 1514.958 ± 0.051m | 1514.659 ± 0.001m |
| **Otse** | Latitude | 25°0'27.67422" | 25°0'27.67414" | 25°0'27.67416" |
|  | Longitude | 25°42'56.03882" | 25°42'56.03896" | 25°42'56.03894" |

| | Ellipsoidal height (m) | 1513.506 ± 0.046 m | 1513.548 ± 0.052m | 1513.452 ± 0.001m |

*Orthometric height*

The geoidal height at Monalanong and Otse were calculated using the coordinates from above and the geoid calculator. The geoidal height is found to be 22.54 m at Monalanong and 23.20 m at Otse hill. Equation 1 was then used to calculate orthometric height at Monalanong and Otse. All three independent processing methods found Monalanong Hill taller than Otse Hill with greater than 95% confidence. See Fig 6, which plots the measured height of Monalanong above Otse for each processing method, including one- and two-sigma error bounds.

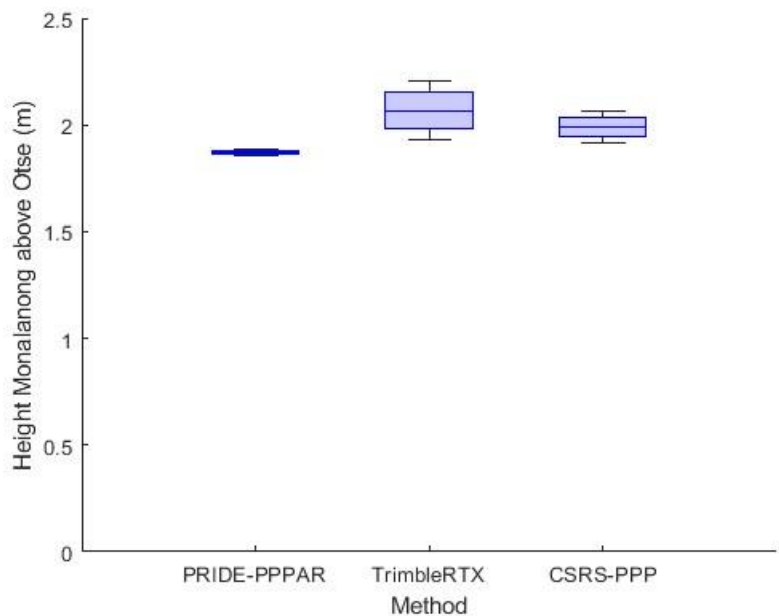

Figure 6: Graph of height difference Monalanong above Otse for each processing method. Boxes are centered at mean values with edges at one-sigma bounds and whiskers at two-sigma 95% confidence interval error bounds.

The PRIDE-PPPAR method resulted in the smallest errors, and the final results will be reported for that method using the static solution (Table 4).

Table 4: Static solution for Monalanong Hill and Otse Hill processed with PRIDE-PPPAR.

| Station | Monalanong | Otse |
|---|---|---|
| Latitude | -24.8404370 | -25.0076873 |
| Longitude | 25.6652302 | 25.7155664 |
| Ellipsoidal Height(m) | 1514.658 | 1513.452 |
| Sigma (vertical, m) | 0.00054 | 0.00071 |
| Orthometric Height (m) | 1492.12 | 1490.25 |

The sigma values for the elevations from static solutions are less than 1 cm. However, because antenna heights were measured to the nearest cm with a tape measure, we will report 95% error bounds as +/-1cm and round orthometric heights to the nearest cm. This means Monalanong Hill was measured to be 1492.12m +/-0.01m and Otse Hill was measured to be 1490.25m +/-0.01m.

**Discussion**

Monalanong Hill (location 24.840437 S, 25.665230 E) is the highest point in Botswana, and is 1.87m +/-0.02m taller than Otse Hill. The error bounds on the elevation measurements do not overlap, meaning Monalanong Hill is higher than Otse hill with greater than 95% confidence.

**Acknowledgements**

The authors would like to thank Matthew Algeo for help in organizing the surveys.

**Declaration of Interest Statement**

The authors have no conflicts of interest to declare.

**Data Availability Statement**

Raw measurement files are available for download at

https://github.com/ericgilbertson1/Botswana

Sampson, S. Kanae & P.D. Bates, A high accuracy map of global terrain elevations, Geophysical Research Letters, vol.44, pp.5844-5853, 2017 doi: 10.1002/2017GL072874.

**Tables**

Table 1: Elevations (m) for each peak from different sources (EGM96 geoid).

Table 2: Duration and number of observations for each location.

Table 3. GNSS post-processing results.

Table 4: Static solution for Monalanong Hill and Otse Hill processed with PRIDE-PPPAR.

**Figures**

Figure 1. Location of Otse Hill and Monalanong Hill in Botswana. Basemap from Opentopo (Opentopo 2025).

Figure 2. Trimble DA2 setup on Monalanong Hill and Otse hill.

Figure 3. Relationship between ellipsoidal height, geoidal height and orthometric height.

Figure 4: Convergence of solutions using CSRS-PPP.

Figure 5: Convergence of solutions using PRIDE –PPP-AR.

Figure 6: Graph of height difference Monalanong above Otse for each processing method. Boxes are centered at mean values with edges at one-sigma bounds and whiskers at two-sigma 95% confidence interval error bounds.